\documentclass[journal=jacsat,manuscript=article]{achemso}

\usepackage[version=3]{mhchem} 

\author{Shigeyuki Ishida}
\affiliation[University of Tokyo]{Department of Physics, University of Tokyo, Tokyo 113-0033, Japan}
\alsoaffiliation[National Institute of Advanced Industrial Science and Technology]{National Institute of Advanced Industrial Science and Technology, Tsukuba 305-8568, Japan}
\alsoaffiliation[Japan Science and Technology Agency]{JST, Transformative Research-Project on Iron Pnictides, Tokyo 102-0075, Japan}
\email{s.ishida@aist.go.jp}
\phone{+81-29-861-2330}
\fax{+81-29-861-5569}

\author{Masamichi Nakajima}
\affiliation[University of Tokyo]{Department of Physics, University of Tokyo, Tokyo 113-0033, Japan}
\alsoaffiliation[National Institute of Advanced Industrial Science and Technology]{National Institute of Advanced Industrial Science and Technology, Tsukuba 305-8568, Japan}
\alsoaffiliation[Japan Science and Technology Agency]{JST, Transformative Research-Project on Iron Pnictides, Tokyo 102-0075, Japan}

\author{Tian Liang}
\affiliation[University of Tokyo]{Department of Physics, University of Tokyo, Tokyo 113-0033, Japan}
\alsoaffiliation[National Institute of Advanced Industrial Science and Technology]{National Institute of Advanced Industrial Science and Technology, Tsukuba 305-8568, Japan}
\alsoaffiliation[Japan Science and Technology Agency]{JST, Transformative Research-Project on Iron Pnictides, Tokyo 102-0075, Japan}

\author{Kunihiro Kihou}
\affiliation[National Institute of Advanced Industrial Science and Technology]{National Institute of Advanced Industrial Science and Technology, Tsukuba 305-8568, Japan}
\alsoaffiliation[Japan Science and Technology Agency]{JST, Transformative Research-Project on Iron Pnictides, Tokyo 102-0075, Japan}

\author{Chul-Ho Lee}
\affiliation[National Institute of Advanced Industrial Science and Technology]{National Institute of Advanced Industrial Science and Technology, Tsukuba 305-8568, Japan}
\alsoaffiliation[Japan Science and Technology Agency]{JST, Transformative Research-Project on Iron Pnictides, Tokyo 102-0075, Japan}

\author{Akira Iyo}
\affiliation[National Institute of Advanced Industrial Science and Technology]{National Institute of Advanced Industrial Science and Technology, Tsukuba 305-8568, Japan}
\alsoaffiliation[Japan Science and Technology Agency]{JST, Transformative Research-Project on Iron Pnictides, Tokyo 102-0075, Japan}

\author{Hiroshi Eisaki}
\affiliation[National Institute of Advanced Industrial Science and Technology]{National Institute of Advanced Industrial Science and Technology, Tsukuba 305-8568, Japan}
\alsoaffiliation[Japan Science and Technology Agency]{JST, Transformative Research-Project on Iron Pnictides, Tokyo 102-0075, Japan}

\author{Teruhisa Kakeshita}
\affiliation[University of Tokyo]{Department of Physics, University of Tokyo, Tokyo 113-0033, Japan}
\alsoaffiliation[Japan Science and Technology Agency]{JST, Transformative Research-Project on Iron Pnictides, Tokyo 102-0075, Japan}

\author{Yasuhide Tomioka}
\affiliation[National Institute of Advanced Industrial Science and Technology]{National Institute of Advanced Industrial Science and Technology, Tsukuba 305-8568, Japan}
\alsoaffiliation[Japan Science and Technology Agency]{JST, Transformative Research-Project on Iron Pnictides, Tokyo 102-0075, Japan}

\author{Toshimitsu Ito}
\affiliation[National Institute of Advanced Industrial Science and Technology]{National Institute of Advanced Industrial Science and Technology, Tsukuba 305-8568, Japan}
\alsoaffiliation[Japan Science and Technology Agency]{JST, Transformative Research-Project on Iron Pnictides, Tokyo 102-0075, Japan}

\author{Shin-ichi Uchida}
\affiliation[University of Tokyo]{Department of Physics, University of Tokyo, Tokyo 113-0033, Japan}
\alsoaffiliation[Japan Science and Technology Agency]{JST, Transformative Research-Project on Iron Pnictides, Tokyo 102-0075, Japan}

\title[]
  {Effect of doping on the magnetostructural ordered phase of iron arsenides: A comparative study of the resistivity anisotropy in the doped BaFe$_2$As$_2$ with doping into three different sites}

\abbreviations{}

\begin{document}
\begin{abstract}
In order to unravel a role of doping in the iron-based superconductors, we investigated the in-plane resistivity for BaFe$_2$As$_2$ doped at either of the three different lattice sites, Ba(Fe$_{1-x}$Co$_x$)$_2$As$_2$, BaFe$_2$(As$_{1-x}$P$_x$)$_2$, and Ba$_{1-x}$K$_x$Fe$_2$As$_2$, focusing on the doping effect in the low-temperature antiferromagnetic/orthorhombic (AFO) phase. A major role of doping in the high-temperature paramagnetic/tetragonal (PT) phase is known to change the Fermi surface by supplying charge carriers or by exerting chemical pressure. In the AFO phase, we found a clear correlation between the magnitude of residual resistivity and resistivity anisotropy. This indicates that the resistivity anisotropy originates from the anisotropic impurity scattering from dopant atoms. The magnitude of residual resistivity is also found to be a parameter controlling the suppression rate of AFO ordering temperature $T_s$. Therefore, the dominant role of doping in the AFO phase is to introduce disorder to the system, distinct from that in the PT phase.
\end{abstract}

\section{Introduction}
The iron arsenides, which are in most cases antiferromagnetic (AF) metals with orthorhombic lattice distortions in their parent phase, can be turned into high-transition-temperature (high-$T_c$) superconductors by chemical substitution/doping~\cite{Kamihara,Rotter,Sefat,Jiang}. The temperature ($T$) - doping ($x$) phase diagram of iron pnictides is similar to that of high-$T_c$ cuprates in that the system moves from an AF phase to a superconducting (SC) phase as the doping level increases. As in the case of the high-$T_c$ cuprates, it is worth investigating the evolution of the electronic state with doping in order to understand the real nature of the AF metallic state and how it is linked to the SC phase. 

One notable feature of the antiferromagnetic/orthorhombic (AFO) phase of iron-based superconductors is an anisotropic electronic state. In the AFO phase of iron arsenides, anisotropic electronic state has been revealed by neutron scattering~\cite{Zhao}, transport~\cite{Chu} and optical measurements~\cite{Nakajima2}, angle-resolved photoemission spectroscopy (ARPES)~\cite{Yi1}, scanning tunneling spectroscopy (STS)~\cite{Chuang}. Except for the resistivity anisotropy, the origin of the anisotropy in these spectra is inherent to the electronic state of the AFO phase which is characterized by the stripe-AF spin order and orthorhombic lattice distortions possibly connected to orbital ordering/polarization. Extensive theoretical and experimental efforts have been devoted to understand its origin.

What are the roles played by chemical doping is another important issue. In the case of high-$T_c$ cuprates, a crucial role of doping is to tune the carrier concentration, and in some cases to introduce disorder~\cite{Fukuzumi,Eisaki}. In an analogous way, for example, Co (K) substitution for Fe (Ba) in BaFe$_2$As$_2$ chemically works as electron (hole) doping, which is also supported by the evolution of the volume of Fermi surfaces (FS) observed by ARPES~\cite{Liu,Malaeb}. On the other hand, the phase diagram of BaFe$_2$As$_2$ with P substituted for As, which does not change the carrier concentration (dubbed as isovalent doping), is similar to that in other doping cases~\cite{Jiang} This suggests that a change of the carrier concentration is not an only effect of doping.   

So far, the doping evolution of physical properties has been investigated for one particular system. However, in order to achieve the unified view of the phase diagram of iron-based superconductors, a comprehensive study is necessary to investigate the similarities and differences among various doping routes toward superconducting phase. Here, we focus on the AFO phase of the representative iron arsenide BaFe$_2$As$_2$, and investigated Co-, P-, and K-doping effect by studying the doping evolution of the in-plane resistivity and its anisotropy. It is found that in all the three cases major role of doping is to introduce disorder the effect of which strongly dependent on the dopant site. We show that strength of the scattering from dopant atoms controls the suppression rate of the AFO phase as well as the resistivity anisotropy.  

\section{Results and discussion}

\subsection{Doping evolution of the in-plane resistivity in the AFO phase}

\begin{figure}[t]
\includegraphics[width=0.9\columnwidth,clip]{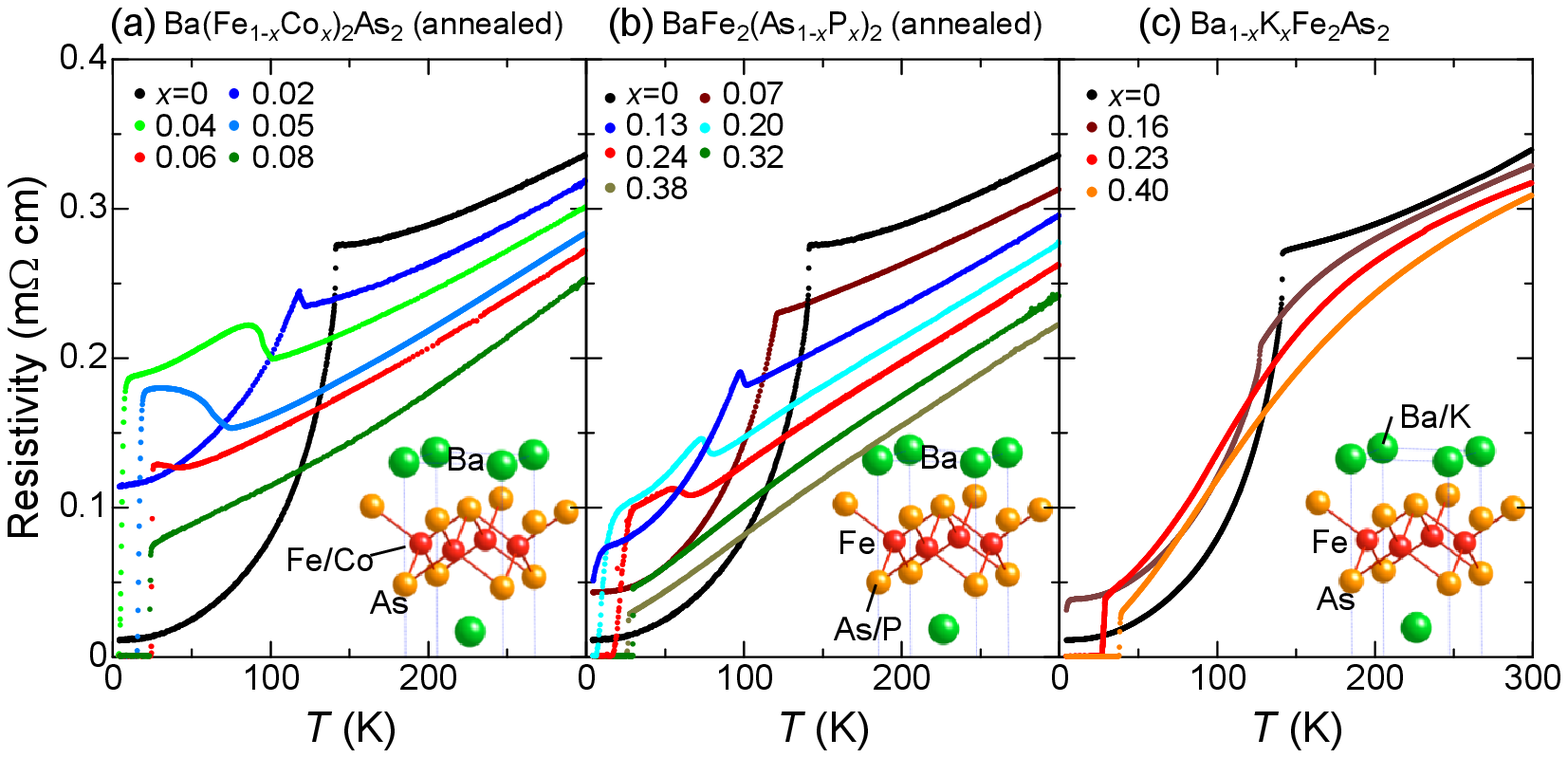}
\caption{\label{fig1} (Color online) Doping evolution of the temperature dependence of the in-plane resistivity for Ba(Fe$_{1-x}$Co$_x$)$_2$As$_2$ (a), BaFe$_2$(As$_{1-x}$P$_x$)$_2$ (b), and Ba$_{1-x}$K$_x$Fe$_2$As$_2$ (c) in the underdoped regime.}
\end{figure}

Temperature dependences of the in-plane resistivity are shown in Figs.~\ref{fig1}(a)-(c) for Ba(Fe$_{1-x}$Co$_x$)$_2$As$_2$, BaFe$_2$(As$_{1-x}$P$_x$)$_2$, and Ba$_{1-x}$K$_x$Fe$_2$As$_2$, covering the doping range from $x$ = 0 to the composition just above the AFO-SC coexisting region. The resistivity of parent BaFe$_2$As$_2$ exhibits an abrupt decrease below $T_s$ = 143~K associated with the transition from the paramagnetic-tetragonal (PT) phase to the AFO phase. The decrease of resistivity despite the loss of the carrier density is due to a reconstruction of Fermi surface in the AFO phase which generates high-mobility carriers dominating the charge transport~\cite{Ishida,Terashima}. The residual resistivity is quite low ($\rho_0$ $\sim$~10~$\mu\Omega$cm) for a well annealed high-quality crystal. 

When Co is substituted for Fe, the residual resistivity in the AFO phase rapidly increases up to $x$~=~0.04 (see Fig.~\ref{fig2}(a)), which indicates that the Co atom works as a strong scattering center. For $x$~=~0.02, $\rho_0$/$x$ $\sim$~60~$\mu\Omega$cm/$x$($\%$Co), which is comparable with the residual resistivity produced by a Zn impurity introduced into the underdoped cuprates~\cite{Fukuzumi}. With further doping, however, residual resistivity starts to decrease accompanied by the appearance of superconductivity. A drop in the scattering rate is observed in the infrared spectrum for Ba(Fe$_{1-x}$Co$_x$)$_2$As$_2$ for $x$'s beyond 0.04.~\cite{Nakajima1} This result is suggestive of a formation of unusual impurity states around a Co atom in the AFO phase. 

In the case of BaFe$_2$(As$_{1-x}$P$_x$)$_2$, the doping evolution of the resistivity is qualitatively similar to that of Ba(Fe$_{1-x}$Co$_x$)$_2$As$_2$. However, the magnitude of residual resistivity (Fig.~\ref{fig2}(b)) is by an order smaller than that of Ba(Fe$_{1-x}$Co$_x$)$_2$As$_2$, indicating that a P atom substituted for As scatters carriers less strongly than Co for Fe site. The isovalency of P to As is probably responsible for the weaker scattering. 

The doping evolution of the in-plane resistivity of Ba$_{1-x}$K$_x$Fe$_2$As$_2$ is entirely different from the above two cases. The residual resistivity is by an order of magnitude smaller. Considering that annealing is difficult for Ba$_{1-x}$K$_x$Fe$_2$As$_2$ crystals and as-grown crystals likely contain crystal disorder/deficiency near the FeAs block, the K-induced residual resistivity of Ba$_{1-x}$K$_x$Fe$_2$As$_2$ would be practically zero over the whole doping range. Therefore, the magnitude of the residual resistivity or the strength of the impurity scattering by dopant atoms gets smaller as the dopant site is farther away from the Fe plane. Note that the scattering strength of individual dopant atom shows an overall decrease with increasing dopant concentration, certainly associated with weakening of the AFO order~\cite{Nakajima3}. Below, we show that the strength of impurity scattering has an intimate relation with the magnitude of the resistivity anisotropy as well as the suppression rate of the AFO order. 

\subsection{Suppression of AFO phase by chemical substitution}

In the case of cuprates, the AF order is rapidly destroyed by doping small amount of $\sim$~2$\%$ holes~\cite{Keimer} but it is robust to the Zn substitution for Cu and persists at Zn concentration as large as 25$\%$~\cite{Hucker}. These experimental results can be reproduced by the magnetic frustration and magnetic dilution model for 2-dimensional Heisenberg antiferromagnet~\cite{Korenblit}, respectively. We investigate the suppression rate (- d$T_s$/d$x$) of the AFO order for three kinds of doping into BaFe$_2$As$_2$. 

\begin{figure}[t!]
\includegraphics[width=0.8\columnwidth,clip]{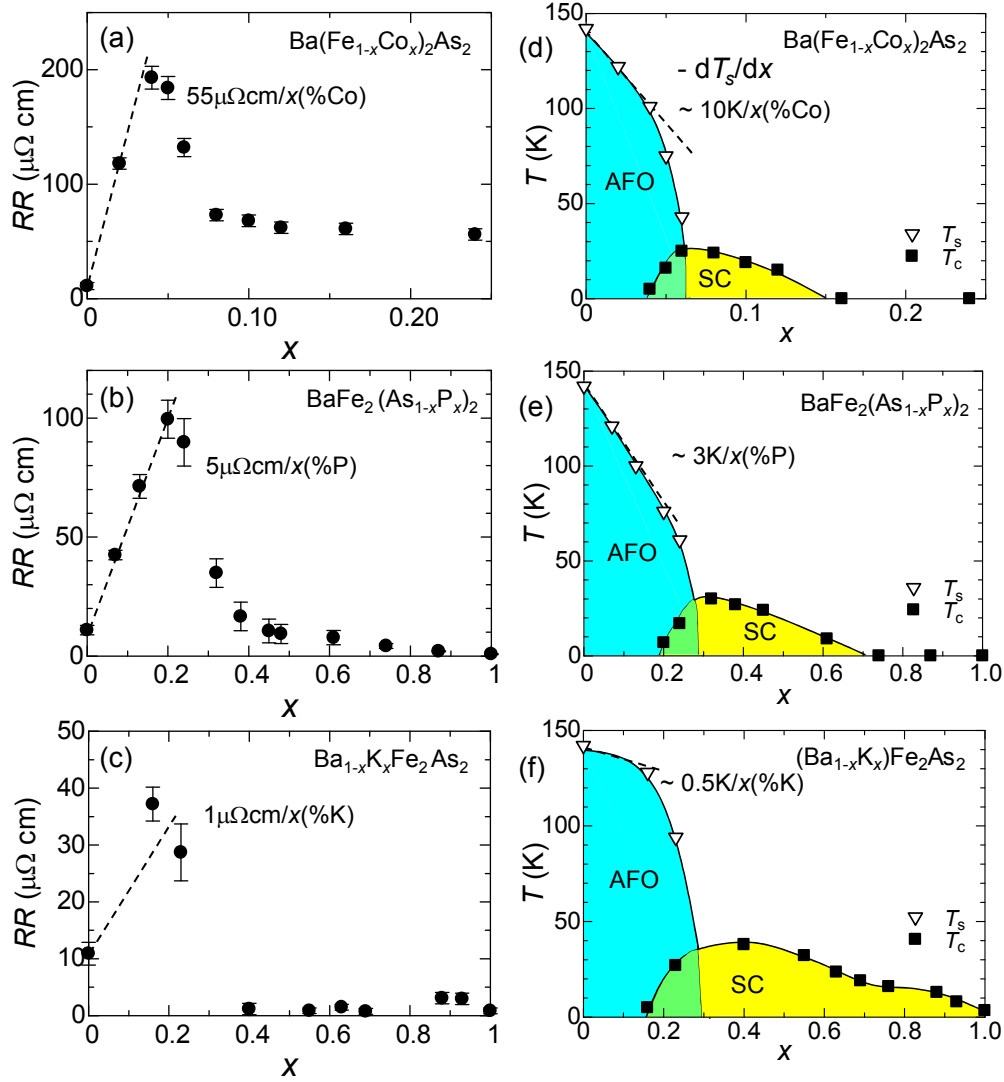}
\caption{\label{fig2} (Color online) Doping dependence of residual resistivityof Ba(Fe$_{1-x}$Co$_x$)$_2$As$_2$ (a), BaFe$_2$(As$_{1-x}$P$_x$)$_2$ (b), and Ba$_{1-x}$K$_x$Fe$_2$As$_2$ (c). Dashed lines indicate the increase of residual resistivity upon doping and the increasing rates (slope for each compound) are also shown. (d)-(f) Phase diagram for each compound determined based on the in-plane resistivity measurement. There is also shown the suppression rate of AFO order (- d$T_s$/d$x$).}
\end{figure}

In the case of Co doping, the AFO phase is radically suppressed with a rate of (- d$T_s$/d$x$) $\sim$ 10K/$x$($\%$Co) below $x$ = 0.05 (Fig.~\ref{fig2}(d)). With further doping, the superconductivity appears ($x$ $\geq$ 0.05) and the decrease of $T_s$ speeds up. If one assumes that each Co atom introduces one mobile electron, though there are controversial views on whether or not the substituted Co atoms supply carriers~\cite{Levy,Wadati,Kemper}, it seems to frustrate the Fe spin order. 

The isovalent P substitution neither introduces charge carrier nor dilutes the Fe spins, so the effect of P substitution on the AF spin order is expected to be very weak. However, as in the case of Co doping, P doping reduces $T_s$ with a rate of (- d$T_s$/d$x$) $\sim$~3K/$x$($\%$P) for $x$ $<$ 0.2, about one-third of that for Co doped case (Fig.~\ref{fig2}(e)). Unusual is the K doping case. A K atom substituted for Ba adds one hole to the AF order state, but the suppression rate of $T_s$ is very small at least in the low doping region (Fig.~\ref{fig2}(f)). We have only three data points in underdoped regime, but from the data in Refs.~\cite{Avci,Ohgushi} it is confirmed that the suppression of $T_s$ is very slow with (- d$T_s$/d$x$) $<$~0.5K/$x$($\%$K) up to $x$ $<$~0.15, although $T_s$ starts to rapidly drop once superconducting phase appears ($x$ $>$ 0.15) and coexists with the AFO phase.

Considering these results for the three-types of doping, the mechanism of suppression of the AFO phase with doping in BaFe$_2$As$_2$ is quite different from that of the localized spin AF order in cuprates. Interestingly, we notice that for the three types of doping the suppression rate of the AFO phase decreases in the order of decreasing residual resistivity produced by doping. This is suggestive of disorder effect playing a substantial role in the suppression of the AFO phase. Vavilov and Chubukov, based on the itinerant magnetism (spin density wave (SDW)), explain a linear suppression of $T_s$ ($T_{\rm N}$) as a disorder/impurity effect with the suppression rate determined by the carrier scattering rate~\cite{Vavilov}, in apparent agreement with the present result. In fact, the $T_s$ suppression rate is also very small in the 1111 system~\cite{Luetkens,Hess} ($Ln$FeAsO$_{1-x}$F$_{x}$ and $Ln$FeAsO$_{1-y}$, $Ln$ being rare-earth element) where the dopant atoms or vacancies are located at the O site far away from the FeAs block. 

At this point, one should also consider a possibility that a change of lattice parameters with doping is another driving force to suppress the AFO order. It is known that the AFO order is also suppressed by the application of hydrostatic pressure~\cite{Colombier}. For P doping the lattice constants appreciably shrink in both $a$ and $c$ directions, so the reduction in volume is most remarkable~\cite{Kasahara}. The relatively rapid decrease of $T_s$ in the P-doping case despite that the residual resistivity is by a factor of $\sim$~10 smaller than that of Co-doped compound, indicates that the chemical pressure effect is also at work in suppressing the AFO order. Considering that the volume reduction rate is smaller for Co doping\cite{Ni} and much smaller for K doping~\cite{Rotter2}, the chemical pressure effect is not expected to be significant in these two cases.

We see that in all the three cases $T_s$ rapidly decreases once the superconducting phase appears and coexists with the AFO order. A naive explanation for this is that the AFO-PT/SC transition is essentially of 1st order, and hence $T_s$ should have discontinuously dropped at the transition. In this context a small spatial fluctuation of dopant concentration would lead to the coexistence/phase-separation of the two phases, and make $T_s$ decrease continuously but rapidly. As a supportive evidence, in the case of application of pressure, which is a cleaner control parameter than chemical doping, and in the doped 1111 system where the effect of dopant disorder is weakest, $T_s$ sharply drops and there is almost no AFO-SC coexisting region~\cite{Colombier}.  

It should be noted that a major effect of these chemical dopings in the PT phase is to change the Fermi surface by adding extra electrons or holes or by exerting chemical pressure~\cite{Malaeb,Liu2,Yoshida}. Disorder effect due to dopant impurities may be seen in the superconducting dome ($T_c$-$x$ curve). The maximum value of $T_c$ and the width of the dome decrease in the order of K, P, and Co doping (see Fig.~\ref{fig2}(d)-(f)), just in the order of increasing disorder strength observed in the AFO phase.

\subsection{Origin of the in-plane resistivity anisotropy}

Originally, the resistivity anisotropy was considered to arise directly from the intrinsically anisotropic electronic state of the AFO phase. For the Co-doped system, the resistivity along the shorter $b$ axis with ferromagnetic spin alignment ($\rho_b$) is always higher than that along the longer $a$ axis with antiferromagnetic spin alignment ($\rho_a$). This looks odd in view of, e.g., the double-exchange mechanism, but is in agreement with the anisotropy in the low-energy optical conductivity ($\sigma_a$ $>$ $\sigma_b$)~\cite{Nakajima2}. The anisotropy in optical conductivity is explained by the theories taking in to account of stripe AF spin order and/or orbital correlations (ordering)~\cite{Yin}.

We investigated, in the preceding work, the anisotropy of the in-plane resistivity in the AFO phase of underdoped Ba(Fe$_{1-x}$Co$_x$)$_2$As$_2$~\cite{Ishida2}. What we found are (1) the resistivity anisotropy at low temperatures almost vanishes for clean BaFe$_2$As$_2$, (2) a finite anisotropy is induced by Co doping in the residual resistivity component, and (3) the anisotropy in as-grown crystals arises probably from the crystal defects present nearby FeAs blocks. These findings evidence that the resistivity anisotropy originates from the anisotropic impurity scattering by doped Co atoms / crystal defects. A Co impurity atom introduced into the AFO phase is supposed to polarize its electronic surrounding with intrinsic anisotropy, and thereby it works as an anisotropic scattering center. The impurity-induced resistivity anisotropy scenario is supported by the optical measurement performed on the detwinned Ba(Fe$_{1-x}$Co$_x$)$_2$As$_2$~\cite{Nakajima3}. The width of the Drude component, which is proportional to the carrier scattering rate, 1/$\tau$, is found to increase in proportion to the Co concentration and become larger along $b$ axis than along $a$ axis. Moreover, the recent scanning-tunneling-spectroscopy (STS) measurement has discovered the formation of $a$-axis aligned electronic dimers surrounding each Co in the AFO state of Ca(Fe$_{1-x}$Co$_x$)$_2$As$_2$~\cite{Allan}, in agreement with our speculation of the anisotropic Co impurity state. 

\begin{figure}[t!]
\includegraphics[width=0.9\columnwidth,clip]{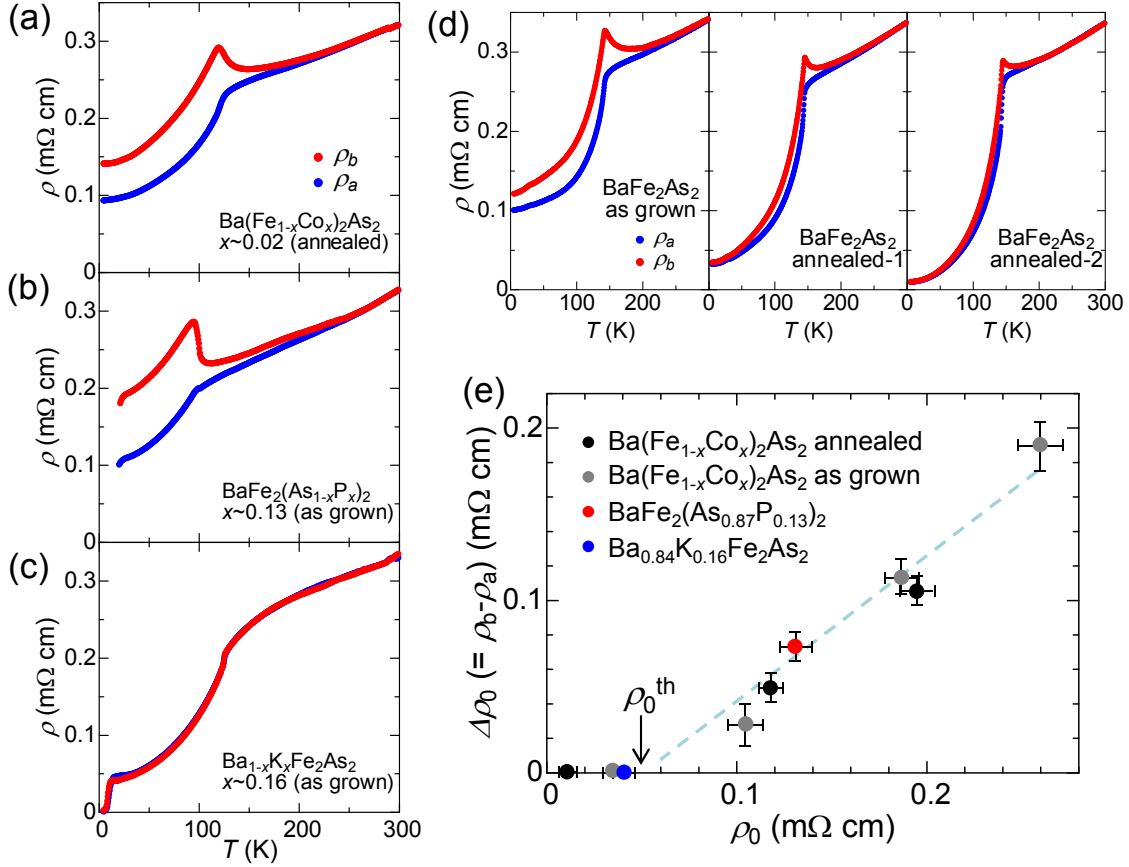}
\caption{\label{fig3} (Color online) The in-plane resistivity anisotropy of selected compounds, annealed Ba(Fe$_{0.98}$Co$_{0.02}$)$_2$As$_2$ (a), as-grown BaFe$_2$(As$_{0.87}$P$_{0.13}$)$_2$ (b), as-grown Ba$_{0.84}$K$_{0.16}$Fe$_2$As$_2$ (c) and as-grown and annealed BaFe$_2$As$_2$ (d). Annealing time is different for annealed-1 (shorter) and annealed-2 (longer) BaFe$_2$As$_2$ samples. (e) The difference of residual component of in-plane resistivity ($\Delta\rho_0$ = $\rho_b$ - $\rho_a$) at low temperature against the residual resistivity. $\rho_0^{\rm th}$ indicates the threshold value of residual resistivity where the resistivity anisotropy appears.}
\end{figure}

Here, we extend the measurement to P- and K-doped compounds. Figure~\ref{fig3}(b) and (c) shows the in-plane resistivity anisotropy of BaFe$_2$(As$_{1-x}$P$_x$)$_2$ ($x$ $\sim$~0.13) and Ba$_{1-x}$K$_x$Fe$_2$As$_2$ ($x$ $\sim$~0.16), respectively, together with the result for Ba(Fe$_{1-x}$Co$_x$)$_2$As$_2$ ($x$ = 0.02) which have similar values of $T_{s}$. Unfortunately, the measurements on the P- and K-doped compounds were performed on as-grown crystals. We need to remove damaged or contaminated surface layers of an annealed crystal in order to obtain reliable data. However, since as-grown P- and K-doped crystals were thinner than Co-doped one, the crystals became too thin to apply uniaxial pressure after removing damaged surface layers. 

BaFe$_2$(As$_{1-x}$P$_x$)$_2$ is found to also show a sizable resistivity anisotropy. As in the case of Ba(Fe$_{1-x}$Co$_x$)$_2$As$_2$, the resistivity along the $b$-axis is higher than that along the $a$-axis. However, considering that the dopant concentration $x$ is larger for P-doped case, the magnitude of the anisotropy is smaller. Then, it is likely that a doped P atom forms similar anisotropic impurity state, and acts as an anisotropic scattering center with smaller scattering cross-section than that of Co impurity. 

By contrast, Ba$_{1-x}$K$_x$Fe$_2$As$_2$ does not show a discernible anisotropy (Fig.~\ref{fig3}(c)), in agreement with the previous report~\cite{J.J.Ying}. It was argued that the resistivity anisotropy was a property inherent to electron-doped compounds. However, the presence of anisotropy in the isovalent P-doped compound questions this hypothesis. 

The absence of resistivity anisotropy seems odd in view of the results for Co and P doping, as a sizable residual resistivity (30 - 40~$\mu\Omega$cm) is observed for the K-doped compound. However, collecting the data of the in-plane resistivity anisotropy for various samples including as-grown Ba(Fe$_{1-x}$Co$_x$)$_2$As$_2$, we find a correlation between the magnitude of resistivity anisotropy $\Delta\rho_0$ = $\rho_b$ - $\rho_a$ and the residual resistivity $\rho_0$ measured for free-standing (twinned) crystal which coinsides with the average residual resistivity in ($\rho_a$ + $\rho_b$)/2 (Fig.~\ref{fig3}(e)). Since the magnitude of $\rho_0$ is a measure of the impurity scattering rate, this correlation confirms again our conclusion that the resistivity anisotropy originates from by the impurity scattering, irrespective of the sign of the introduced charge carrier (electron or hole). Furthermore, there appears to be a threshold value of residual resistivity ($\rho_0^{\rm th}$ $\sim$ 50~$\mu\Omega$cm) above which finite $\Delta\rho_0$ appears. This suggests that, when the impurity potential is too weak, the unusual anisotropic impurity state is not formed, as might be the case with K-doping and annealed parent compound.  

As we demonstrated before~\cite{Liang,Ishida}, both residual resistivity and resistivity anisotropy decrease after annealing. Particularly for BaFe$_2$As$_2$, sufficient annealing makes the resistivity anisotropy vanishingly small at low temperatures as shown in Fig.~\ref{fig3}(d). This, in turn, implies that the as-grown crystal might contain defects and impurities nearby FeAs blocks and they also act as an anisotropic scattering center. Then, the resistivity anisotropy for the as-grown BaFe$_2$(As$_{1-x}$P$_x$)$_2$ crystal might be induced by both P atoms and crystal defects. To estimate genuine resistivity anisotropy induced by P atoms, we compared the magnitude of residual resistivity $\rho_0$ between as-grown and annealed BaFe$_2$(As$_{1-x}$P$_x$)$_2$ ($x$ $\sim$~0.13). The values of $\rho_0$ is about 130~$\mu\Omega$cm for the as-grown crystal, which is reduced to 80~$\mu\Omega$cm after annealing. The annealing is expected to considerably reduce the defect density, so $\rho_0$ for the annealed crystal is attributable, in most part to the contribution from the P impurity. Since this value is well above the threshold $\rho_0^{\rm th}$, the P impurity likely induces the residual resistivity. Considering the correlation between $\rho_0$ and $\Delta\rho$ displayed in Fig.~\ref{fig3}(e), $\Delta\rho$ induced by P impurity would be in between 20 to 40~$\mu\Omega$cm. 

As-grown Ba$_{1-x}$K$_x$Fe$_2$As$_2$ is also expected to contain crystal defects. However, in view of the small value of $\rho_0$ ($\sim$ 37~$\mu\Omega$cm) for the $x$ = 0.16 crystal which is below $\rho_0^{\rm th}$, the contribution to $\Delta\rho_0$ from the crystal defects is negligibly small. Probably the formation of crystal defects is inhibited under the condition for the crystal growth of K-doped compounds. Therefore it is reasonable to conclude that the K impurity potential is too weak to induce resistivity anisotropy. A possibility is ruled out that the anisotropy induced by crystal defects accidentally compensate that due K impurity with opposite sign~\cite{Blomberg,Fernandes}. 

\begin{figure}[t!]
\includegraphics[width=0.5\columnwidth,clip]{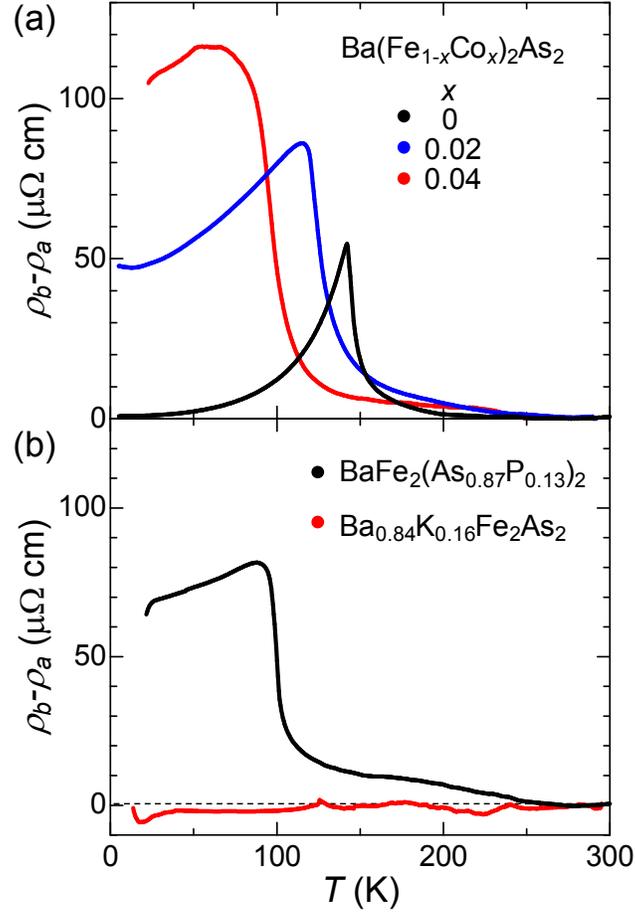}
\caption{\label{fig4} (Color online) Anisotropic resistivity $\Delta\rho$ = $\rho_b$ - $\rho_a$ plotted as a function of temperature for (a) Ba(Fe$_{1-x}$Co$_{x}$)$_2$As$_2$ with $x$ = 0, 0.02 and 0.04 and (b) BaFe$_2$(As$_{1-x}$P$_{x}$)$_2$ with $x$ = 0.13 and Ba$_{1-x}$K$_{x}$Fe$_2$As$_2$ with $x$ = 0.16.}
\end{figure}

We have shown that the anisotropic elastic scattering from dopant impurities is responsible for the resistivity anisotropy at low temperatures in the AFO phase. However, as displayed in Fig.~\ref{fig4}, the anisotropy $\Delta\rho$ = $\rho_b$ - $\rho_a$ gradually increases with raising temperature toward $T_s$. Above $T_s$ $\Delta\rho$ sharply drops after showing a cusp at $T_s$, but remains finite at temperature well above $T_s$. As temperature rises, inelastic scattering processes progressively dominate. So, the enhanced anisotropy at high temperatures is indicative of the presence of anisotropic contribution in the inelastic scattering process. Fernandes $et~al.$ gives rise to the resistivity anisotropy combined with impurity scattering~\cite{Fernandes}. This mechanism might explain the enhanced anisotropy at elevated temperatures below $T_s$ as well as the anisotropy above $T_s$. The anisotropy above $T_s$ (in the tetragonal phase) is usually attributed to a manifestation of rotational symmetry broken nematic phase~\cite{Chu}. However, considering that the resistivity anisotropy arises predominantly from impurity scattering and that the temperature range of the anisotropy above $T_s$ expands with Co/P doping and shrinks after annealing, it would be possible to suppose an extrinsic mechanism, such as short-range AFO order locally induced around impurity atoms, also at work.

\section{Conclusions}
We have investigated the doping evolution and dopant-site dependence of the in-plane resistivity for BaFe$_2$As$_2$ to pursue the effect of doping on the AFO phase. The strength of the dopant impurity scattering is found to be a control parameter of the suppression of the AFO ordering temperature in addition to the chemical pressure exerted by substituted dopant atoms. It also controls the magnitude of the resistivity anisotropy which arises from an anomalous impurity state formed around a dopant atom. The anisotropy diminishes when the dopant site is away from the Fe plane and the impurity potential is too weak to form such an impurity state. Therefore, the major effect of doping on the AFO phase is to introduce disorder. Formation of the anomalous impurity state around a dopant atom is a hallmark of a unique electronic state of the AFO phase which might be regarded as a spin-charge-orbital complex. The fact that the superconducting $T_c$ attains maximum values near the AFO-SC phase boundary suggests that fluctuations of such complex may be relevant to the formation of Cooper pairs in the doped iron arsenides.

\section{Experimental}

Single crystals of Ba(Fe$_{1-x}$Co$_x$)$_2$As$_2$, BaFe$_2$(As$_{1-x}$P$_x$)$_2$, and Ba$_{1-x}$K$_x$Fe$_2$As$_2$ were grown by the self-flux method~\cite{Nakajima1,Nakajima4,Kihou}. The actual compositions of the samples were determined by inductively coupled plasma (ICP) analysis and by the energy dispersive X-ray (EDX) analysis. The crystals were cut in a rectangular shape along the tetragonal [110] directions which become $a$ or $b$ axes in the orthorhombic phase. Typical dimension of the Ba(Fe$_{1-x}$Co$_x$)$_2$As$_2$ crystal is 1.5 $\times$ 1.5~mm$^2$ in the $ab$-plane area and 0.5~mm in thickness along the $c$ axis. In the case of BaFe$_2$(As$_{1-x}$P$_x$)$_2$ and Ba$_{1-x}$K$_x$Fe$_2$As$_2$, the crystals are thinner with 0.1 - 0.2~mm in thickness. The crystals of Co and P doped BaFe$_2$As$_2$ were sealed into an evacuated quartz tube together with BaAs powders and annealed for several days, since annealing remarkably improves transport properties in the ordered phase of BaFe$_2$As$_2$~\cite{Rotundu}, which indicates that the as-grown crystals contain appreciable amount of defects/impurities, and hence the observation of the intrinsic charge transport in this system might be inhibited. However, for K-doped BaFe$_2$As$_2$ annealing damages crystals, so measurements were done on as-grown crystals. For detwinning, the rectangular-shaped crystals were set in an uniaxial pressure cell and detwinned by applying compressive pressure along the tetragonal (110) direction~\cite{Liang}. The resistivity along the $a$ and $b$ axis were measured simultaneously using Montgomery method~\cite{Montgomery} without releasing pressure. The measurements were performed in a Quantum Design physical property measurement system (PPMS). 

\begin{acknowledgement}

SI and MN thank the Japan Society for the Promotion of Science (JSPS) for the financial support. This work was supported by Transformative Research-Project on Iron Pnictides (TRIP) from the Japan Science and Technology Agency, and by the Japan-China-Korea A3 Foresight Program from JSPS, and a Grant-in-Aid of Scientific Research from the Ministry of Education, Culture, Sports, Science, and Technology in Japan.

\end{acknowledgement}




\newpage

\end{document}